\documentstyle[psfig]{l-aa}

\begin{document}

\title {Do the central engines of quasars evolve by accretion ?}

\author{R. Srianand$^1$ and Gopal--Krishna$^2$}

\institute{$^1$Inter-university Centre for Astrophysics \& Astronomy, Pune
University Campus, Pune 411 007, India \\e-mail: anand@iucaa.ernet.in \\
$^2$National Centre for Radio Astrophysics, TIFR,
Poona University Campus\\
Post Bag No.\ 3, Ganeshkhind, Pune 411 007, India \\e-mail:
krishna@ncra.tifr.res.in}

\date{ }
\offprints{R. Srianand}
\thesaurus{   %
              02.12.3 
              11.01.2  
              11.14.1 
              11.17.2 
              11.17.3 
              13.18.1} 
\maketitle
\markboth{}{}

\begin{abstract}

According to a currently popular paradigm, nuclear activity in quasars
is sustained {\it via } accretion of material onto super-massive black holes
located at the quasar nuclei. A useful tracer of the gravitational
field in the vicinity of such central black holes is available in the
form of extremely dense gas clouds within the broad emission-line
region (BLR) on the scale of $\sim 1~$parsec. Likewise, the radio sizes
of the lobe-dominated radio sources are believed to provide a useful
statistical indicator of their ages.  Using two homogeneously observed
(and processed) sets of lobe-dominated radio-loud quasars, taken from
literature, we show that a positive correlation exists between the
radio sizes of the quasars and the widths of their broad $H\beta$
emission lines, and this correlation is found to be significantly
stronger than the other well known correlations involving radio size.
This statistical correlation is shown to be
consistent with the largest (and, hence, very possibly the oldest)
radio sources harboring typically an order-of-magnitude more massive
central engines, as compared to the physically smaller and,
hence, probably much younger radio sources. This inference
is basically in accord with the "accreting central engine" picture for
the radio-loud quasars.


\end{abstract}

\section{Introduction}

Being too compact to be resolved with even the most advanced optical
telescopes, the structure and kinematics of the broad emission line
region (BLR), a prime feature of quasars, continues to be a major
enigma in the AGN research (see, e.g., Brotherton, 1996; Marziani et al.,
1996; Corbin, 1992).  Nonetheless, it is a key ingredient to the
theoretical models that seek to explain various observable properties
of quasars, e.g., their intense $\gamma$-ray emission (e.g., Dermer \&
Schlickeiser, 1993; Ghisellini \& Madau, 1996).  Deciphering the BLR geometry
has therefore been a
key objective of many observational programmes.  One strategy
for this is the so
called `reverberation mapping' (e.g., Peterson, 1993; Koratkar \&
Gaskell, 1991), from which a positive dependence of the BLR size on the
bolometric luminosity has been inferred:  $r~\sim 0.06 L_{46}^{0.5}~pc$
(Netzer \& Laor 1993), where $L_{46}$ is the luminosity expressed in the units
of $10^{46} erg.s^{-1}$.
\begin{table*}
\label{ The sample of lobe dominated radio-loud quasars (48 quasars)}
\begin{tabular}{lrrcccrrc}
\multicolumn {9}{c}{\bf Table 1: Sample of lobe-dominated quasars (42
quasars)}\\
\multicolumn {9}{c}{}\\
\hline
\hline
\multicolumn {1}{c}{QSO}&\multicolumn {1} {c}{z} &
\multicolumn {1}{c} {LAS}& \multicolumn {1}{c}{log($f_c$)}&
\multicolumn {1}{c} {$\rm log(R_v)$}&\multicolumn {1}{c}
{$\rm M_{v}$}&\multicolumn {2}{c}{W($\rm H\beta$)
(km s$^{-1}$)}&\multicolumn {1}{c}{ $\xi$}\\
\cline {7-8}
&&\multicolumn {1}{c} {(arcsec)}&&&&B(96)&JB(91)&\\
\hline
0003+158  &0.450&36.0(1)&$-$0.38(a)&1.95&-26.0&4760&...&...\\
0042+101  &0.583&59.0(3)&$-$0.50(b)&1.79&-24.8&...&17774&...\\
0044+030  &0.624&18.6(5)&$-$0.42(a)&0.83&-27.2&5100&...&...\\
0110+297  &0.363&76.2(2)&$-$0.80(b)&1.59&-24.8&...&8702&...\\
0115+027  &0.670&13.1(2)&$-$0.50(b)& 2.25&-26.5&5000&7591&0.66\\
0118+034  &0.765&45.0(4)&$-$1.20(b)&2.17&-25.1&...&22403&...\\
0133+207  &0.425&68.0(2)&$-$1.00(b)&2.51&-24.0&...&17403&...\\
0134+329  &0.367&1.3(3)&$-$1.14(a)&2.50&-25.7&3800&5863&0.65\\
0405$-$123&0.575&31.7(6)&$-$0.23(a)&2.36&-28.4&4800&...&...\\
0414$-$060&0.781&36.4(4)&$-$0.50(a)&1.68&-27.8&8200&16602&0.49\\
0518+165  &0.759&0.8(9)&$-$0.90(b)&3.94&-24.1&...&4876&...\\
0538+498  &0.545&0.2(7)&$-$1.40(b)&3.43&-24.2&...&3456&...\\
0710+118  &0.768&48.0(2)&$-$1.85(a)&1.09&-27.1&20000&17774&1.13\\
0800+608  &0.689&25.0(10)&$-$0.60(b)&2.65&-24.2&...&6912&...\\
0837$-$120&0.200&169.0(2)&$-$0.70(a)&1.80&-24.7&6060&...&...\\
0838+133  &0.680&10.0(7)&$-$0.50(b)&3.16&-25.4&3000&4197&0.71\\
0903+169  &0.410&50.0(8)&$-$1.40(c)&1.70&-24.5&4400&...&...\\
0952+097  &0.298&12.5(2)&$<$$-$0.50(c)&$<$1.90&-24.1&3800&...&...\\
0955+326  &0.530 &1.0(3)&$-$0.07(a)&2.23&-27.1&1380&...&...\\
1004+130  &0.240&115.0(2)&$-$1.68(a)&0.43&-25.7&6300&9998&0.63\\
1007+417  &0.613&32.0(2)&$-$0.39(a)&2.17&-27.0&3560&6912&0.52\\
1048$-$090&0.334&83.0(2)&$-$1.31(a)&1.66&-24.9&5620&...&...\\
1100+772  &0.311&30.0(1)&$-$0.91(a)&1.63&-25.8&6160&7961&0.77\\
1103$-$006&0.423&21.0(2)&$-$0.14(a)&2.20&-25.8&6560&...&...\\
1111+408&0.730&13.2(2)&$-$1.80(b)&1.89&-25.3&...&9134&...\\
1137+660&0.652&44.2(2)&$-$1.07(a)&1.83&-27.1&6060&8702&0.69\\
1223+252&0.268&67.0(2)&$-$1.59(b)&0.99&-23.8&...&9319 &...\\
1250+568&0.321&1.5(2)&$-$1.62(a)&2.01&-23.6&4560&6295&0.72\\
1305+069&0.599&46.5(4)&$<$$-$1.20(a)&$<$1.52&-26.1& 6440&...&...\\
1351+267&0.310&190.0(2)&$-$0.27(a)&1.72&-24.3&8600&...&...\\
1425+267&0.366&230.0(1)&$-$0.40(a)&1.03&-26.2&9410&...&...\\
1458+718&0.905&2.1(2)&$-$1.00(d)&2.74&-27.3&3000 &...&...\\
1512+370&0.371&54.0(1)&$-$0.71(a)&1.88&-25.6&6810&...&...\\
1545+210&0.264&70.0(1)&$-$1.32(a)&1.74&-24.4&7030&...&...\\
1618+177&0.555&48.0(3)&$-$0.67(a)&1.98&-26.5&7000&...&...\\
1622+238&0.927&21.7(2)&$-$1.72(d)&1.63&-26.7&7100&...&...\\
1704+608&0.371&55.0(1)&$-$1.91(a)&0.73&-26.6&6560&...&...\\
1742+617&0.523&40.0(11)& $-$2.00(b)&1.61&-24.0&...&9751&...\\
1828+487&0.691&14.0(11)&$-$0.50(b)&4.05&-26.2&...&9998&...\\
2135$-$147&0.201&150.0(2)&$-$1.13(a)&1.97&-24.9&7300&11479&0.63\\
2251+113&0.323&9.8(2)& $-$1.52(a)&1.00&-25.8&4160&8702&0.48\\
2308+098&0.432&108.0(1)&$-$0.78(a)&1.41&-26.3&7970&...&...\\
\hline
\multicolumn {9}{l}{{\bf References for LAS:} 1 : Kellermann et al. (1994), 2 :
Nilsson et al. (1993), 3 : Singal (1988), }\\
\multicolumn {9}{l}{4 : Kapahi (1995), 5 : Price et al. (1993), 6 :
Morganti et al.(1993), 7 : Bogers et al. (1994),}\\
\multicolumn {9}{l}{ 8 : Bridle et al (1994),9 : Akujor et al. (1993), 10 :
Jackson et al. (1990), 11 : Reid et al. (1995)}\\
\multicolumn {9}{l}{{\bf References for $f_c$:} a : Wills \& Browne (1986),
b : Jackson \& Browne (1991a), }\\
\multicolumn {9}{l}{c : Brotherton (1996), d : Wills et al. (1992), e :
Kellermann et al. (1994)}\\
\end{tabular}
\end{table*}

An early indication about the BLR geometry came from an empirical
study of a heterogeneously selected sample of radio-loud
quasars with the characteristic core-lobe type radio structure.
The study revealed a statistically significant anti-correlation
between the {\it prominence} of the radio core relative to
the lobe, and the FWHM of the
$H\beta$ broad emission line ($W$). It was thus inferred that the BLR
clouds are predominantly confined to a rotating disk-shaped region surrounding
the quasar nucleus and oriented roughly perpendicular to the jet axis
(Wills \& Browne, 1986; hereafter WB86). The significance of this correlation
was found to be considerably lower in a subsequent study, however
(Jackson \& Browne, 1991b). In a recent work, it has been proposed that
the absolute {\it visual} magnitude, $M_v$, of the quasar (plus its host galaxy)
provides a more reliable  measure of the intrinsic power of the central
engine, and therefore the beamed radio core flux normalized by $M_v$ is a better
indicator of the orientation of the jet relative to the line-of-sight
(Wills \& Brotherton, 1995; Brotherton, 1996). Adopting this new
parameter for the core-prominence and employing a larger sample of
quasars , these authors have found a conspicuous anti-correlation
between the core-prominence and the $W(H\beta)$, thereby supporting the
disk-like BLR geometry inferred earlier by WB86 (at least for
the $H\beta$ emitting clouds).

It is now widely believed that the large widths of the BLR emission lines are
a manifestation of the deep gravitational potential at the centers of quasars,
possibly due to super-massive black-holes. The accretion
of the material postulated for sustaining the quasar luminosity is expected to
steadily increase the mass of the central engine
during the lifespan of the nuclear activity.
If this conjecture is basically right, the question arises:  {\sl do we see
in the data any evidence for the postulated growth of the central mass}
(e.g, in the dynamics of the BLR clouds) ? In the present study we shall
examine this issue by employing published
measurements of radio-loud, steep-spectrum quasars.
While the needed reliable estimator of the age is generally
not available for individual sources, the overall radio size can serve as a
useful statistical measure of age (since most radio sources are believed to
steadily grow in size with time; cf. Sect. 3). Thus,
we wish to examine here the nature of any relationship between the
observed linear sizes of steep-spectrum radio quasars and the widths of
their broad H$\beta$ emission lines.

\section{ The datasets of radio-loud quasars}

In order to minimize the projection effects on the measured radio sizes,
we confine our study to lobe-dominated quasars (LDQs).
Measurement of the other parameter, namely, $W$, the FWHM of the broad
H$\beta$ emission line, is rather complicated, due to the contamination
arising from the lines of Fe II and the narrow component of H$\beta$.
Therefore, adopting uniform observing strategy and profile
extraction procedure for the entire sample is an important
pre-requisite for a meaningful interpretation of the data. Using two such
datasets on H$\beta$ line widths available in the literature
(Brotherton, 1996; Jackson \& Browne, 1991), for which we could
also find the requisite radio data,
we investigate the relationship between different radio properties and
W(H$\beta$) (Table 1).  The largest linear sizes,
$l$, of radio emission associated with these quasars are taken from published
radio maps (Table 1);  these correspond to the cosmological
parameters:  $H_o~=~50~$km s$^{-1}$ Mpc$^{-1}$ and $q_o~=~0$, as
adopted throughout this paper.  Details of the two sets of LDQs
(all having $M_v ~\le~-23$ ) are:

\noindent {\bf (a) B(96):} This first set is derived from the sample of
Brotherton (1996), which itself was based on the QSO compilation by
V\'eron-Cetty \& V\'eron (1988). His selection criteria were: (i)
core-fraction, $f_c$ should be available in the literature ($f_c$ is
defined as the ratio of the core to extended flux density, measured
at 5 GHz in the rest-frame of the quasar), (ii) $V \le 18$ mag, (iii)
$z\le 0.95$ and (iv) declination $>-20^{\circ}$. For the selected 60
quasars, optical spectra were taken with a typical spectral resolution
of 2.5 \AA. Out of these, we have selected all the 31 LDQs
(log$f_c<~$0) having an absolute magnitude $M_v$ brighter than $-23$.

\noindent {\bf (b) JB(91):} This set is derived from Jackson \& Browne
(1991a,b), whose sample consists of low-resolution, ($20-25$\AA)
spectra of 53 radio-loud quasars selected using the following criteria:
(i) $log(R)~\ge~1$ [R is defined as the ratio of the flux densities at
6 cm and $4400$ \AA, in the rest-frame], (ii) $V\le$18-mag, (iii) $z\le
0.86$, (iv) right ascension between 16 and 13 hr, and (v) declination
$> -30^{\circ}$. The sample contains 23 LDQs with available H~$\beta$
profile measurements by  Jackson \& Browne (1991b) who determined $W$ using two
different profile extraction procedures. We adopt here the values of
$W$ obtained using the four Gaussian fitting method as it corrects for
the the narrow line component.

As seen from the last column of Table 1, twelve quasars are common to the two
datasets. The ratio of H$\beta$ line-width determined by B(96) to that by
JB(91), designated as $\xi$, has a mean value of 0.68$\pm$0.16. Thus, in most
of the cases, $\xi$ is close to the mean value, the only exception being
the quasar 0710$+$118 (see, Sect. 3). We form a combined dataset by merging
the quasars present exclusively in the set JB(91) with the set B(96),
after multiplying their quoted H$\beta$ widths by the mean value of
$\xi$ found above.
\begin{figure*}
\centerline{\vbox{
\psfig{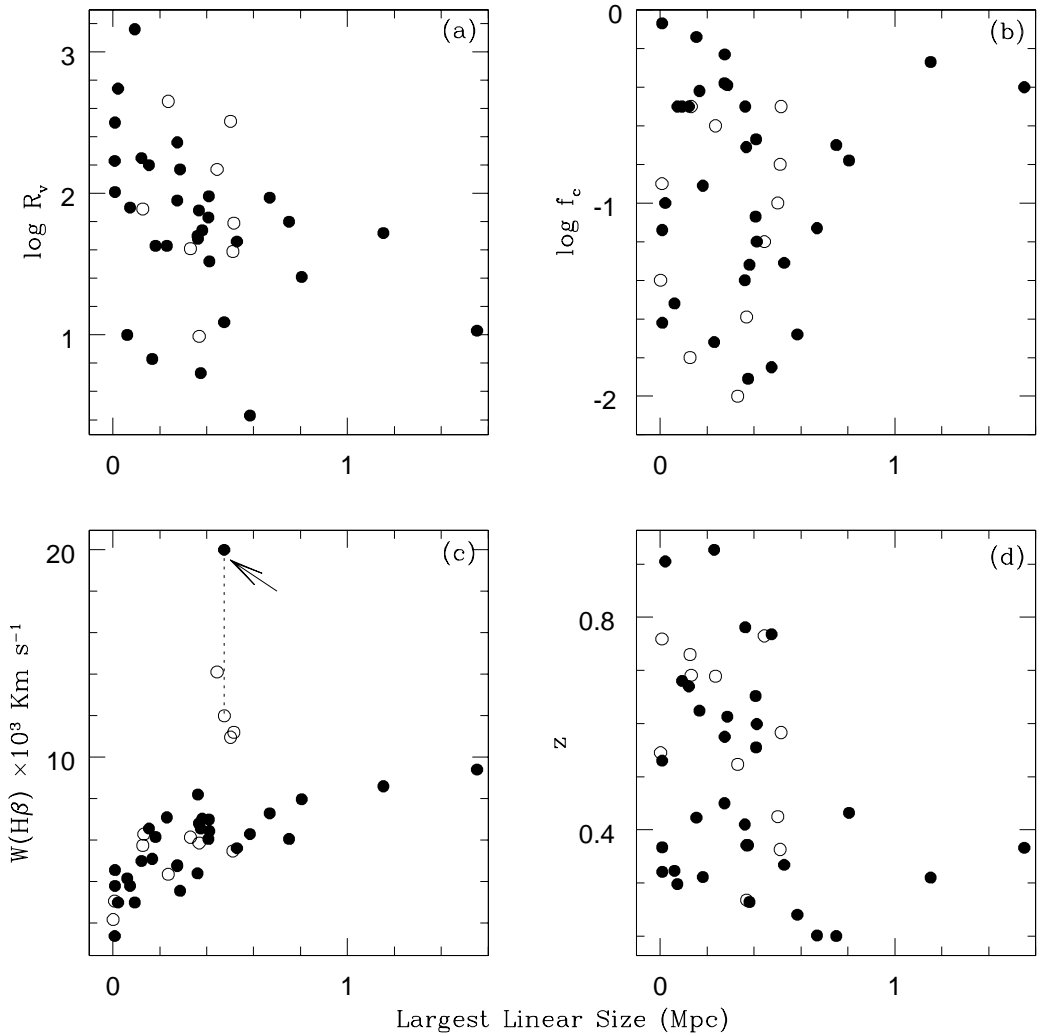}
}}
\caption[]{{\bf (a-d)} The plots of $R_v$, $f_c$, $W(H\beta)$, and $z$,  {\it
versus} the radio size
($l$) for our combined dataset of lobe-dominated quasars. The
filled circle are the data from B(96) and the open circles represent
the data from JB(91) (Sect. 2 and 3).}
\end{figure*}

Table 1 gives a consolidated list of all the LDQs used in our study.
The information provided includes: the IAU name of the quasar,
its redshift (taken from
V\'eron-Cetty \& V\'eron, 1993), the largest angular size (LAS)
measured from the radio maps (references are cited in the footnote).
Column (4) tabulates the value of log $f_c$. Column
(5) gives log $R_v$, defined by Wills \& Brotherton (1995) as $\log R_v
= \log (L_{core}/L_{opt}) = log (L_{core}+M_{v}/2.5) -13.69$, where the
absolute visual magnitude, $M_{v}$, is taken from V\'eron-Cetty \&
V\'eron (1993) and listed in column (6). Columns (7)\&(8) tabulate the
published
widths of the broad H$\beta$ emission line, taken from the two datasets
discussed above. Column (9) lists the value of  $\xi$,
the ratio of W($H\beta$)
measured by B(96) to that by JB(91) for the 12 quasars which are
common to the two data sets.
\begin{table*}
\begin{tabular}{lcrcccrr}
\multicolumn {7}{c}{\bf Table 2: Spearman rank$-$order correlation
coefficients}\\
\\
\hline
\hline
\multicolumn {1} {c} {Sample}& \multicolumn {1} {c} {Size}&
\multicolumn {1} {c} {${l-z}$} &\multicolumn {1}{c}{${l-log(f_c)}$}&
\multicolumn {1} {c} {${l-log(R_v)}$}&
\multicolumn {1} {c} {${l-W}$}&\multicolumn {1} {c} {${log(f_c)-W}$}&
\multicolumn {1} {c} {${log(R_v)-W}$}\\
\hline
B(96) & 31&$-$0.163&$-$0.207&$-$0.376&0.771&$-$0.254&$-$0.397\\
      &&0.373&0.254&0.030&$2.4\times10^{-7}$&0.161&0.024\\
combined&42&$-$0.237&$-$0.111&$-$0.420& 0.765&$-$0.183&$-$0.324\\
dataset&&0.127&0.447&0.005&$2.3\times10^{-9}$&0.240&0.034\\
\hline
\hline
\end{tabular}
\end{table*}

\section {Results}
 
Fig. 1 shows the radio size, $l$, plotted against the parameters, $R_v$,
$f_c$, $W(H\beta)$ and $z$ for our sample of
42 LDQs.  The filled circles are the data from B(96) and the open
circles refer to the data from JB(91). The values of W shown
by the open circles have been rescaled, as discussed
above.  Table 2 shows the results of the non-parametric Spearman rank
correlation tests between the different parameters, for B(96) as well as
the combined dataset. For each case, the upper line gives the
correlation coefficients and the lower line gives the two-sided
significance level of its deviation from zero (smaller value implies a
stronger correlation).  The remarkable trend noticed from Fig. 1 and Table 2 is
that  $l$ is found to correlate more strongly with $W$ than with any of
the other parameters, namely $f_c$, $R_v$, and even redshift, $z$.
The $l-W(H\beta)$ correlation
thus emerges to be of the primary statistical significance
, in agreement with the preliminary results 
reported by Gopal-Krishna (1995) (see, also, Gopal-Krishna \& Srianand, 1998).
It may be recalled
that using a much more limited dataset, Miley \& Miller (1979) had
earlier noticed a positive correlation between $l$ and $W(H\beta)$. However,
almost $\sim40\%$ of their sample was comprised of core-dominated
quasars and hence, the linear sizes used by them are likely to be
influenced by projection effects to a much greater degree (and,
consequently, be rendered less suitable as a measure of source age, {\it
vis-a-vis} the present study where only lobe-dominated quasars
have been considered). Note that the quasar 0710$+$118 (marked with an arrow in
Fig 1c) shows a large deviation from the general trend. Since this quasar 
is the only case where W(H$\beta$) measured by B(96) is larger than that 
found by JB(91) (Table 1), we have also plotted in Fig 1c the scaled  JB(91)
value of its W(H$\beta$) [the open circle connected by
a dotted line] and this point is fully consistent with the general trend.

In order to understand possible implication of the $l - W$ correlation,
we first recall the general consensus that a vast majority of powerful
extragalactic radio sources
grow in linear size with increasing age (e.g., Readhead, 1995; Fanti et
al., 1995).  Thus, to a first order,  the observed linear size, $l$
(i.e., the separation between the outermost peaks in the radio
lobe-pair) can be regarded as a meaningful statistical
indicator of the age (e.g., Best et al., 1996). For individual sources,
radio spectral gradient is also used as a  measure of the age (sect. 4).
However at present such data are largely confined to very extended radio
sources with prominent diffuse lobes, mainly radio galaxies where the
contamination of the lobe emission by the jet component is
usually quite small (e.g., Alexander \& Leahy, 1984; Carilli et al.,
1991;  Liu \& Pooley, 1992; Rawlings \& Saunders, 1991).

A potentially complicating factor in the use of $l$ is the
foreshortening caused due to the (unknown) projection effects. However
our selection
criterion, which excludes core-dominated quasars, would effectively
minimize the role of projection. It may be recalled that $l$ is known
to {\it anti-correlate} with $z$ for quasars, as reported, e.g., by Wardle 
\& Miley (1974), which is usually interpreted in terms of an increase in
the density of the ambient medium. However, this may not play a major
role in our data as it covers a much smaller range in redshift (Table 1).
This expectation is supported by the apparent weakness of the $l - z$ 
correlation in our
data (Table 2). Radio size, $l$, is also known to be anti-correlated with
the radio core prominence, $f_c$, which is 
consistent with the idea of beamed core radiation (e.g., Kapahi \&
Saikia, 1982; Browne \& Perley, 1986; Lister et al., 1994). Again, this trend 
is at best only weakly present in our datasets (Fig. 1; Table 2), which is in
accord with the expectation that projection effects should be of minor
importance in the case of lobe-dominated quasars. This is further supported
by the apparent weakness of $l - R_v$ anti-correlation in our datasets 
(Fig. 1; Table 2).

\begin{figure*}
\psfig{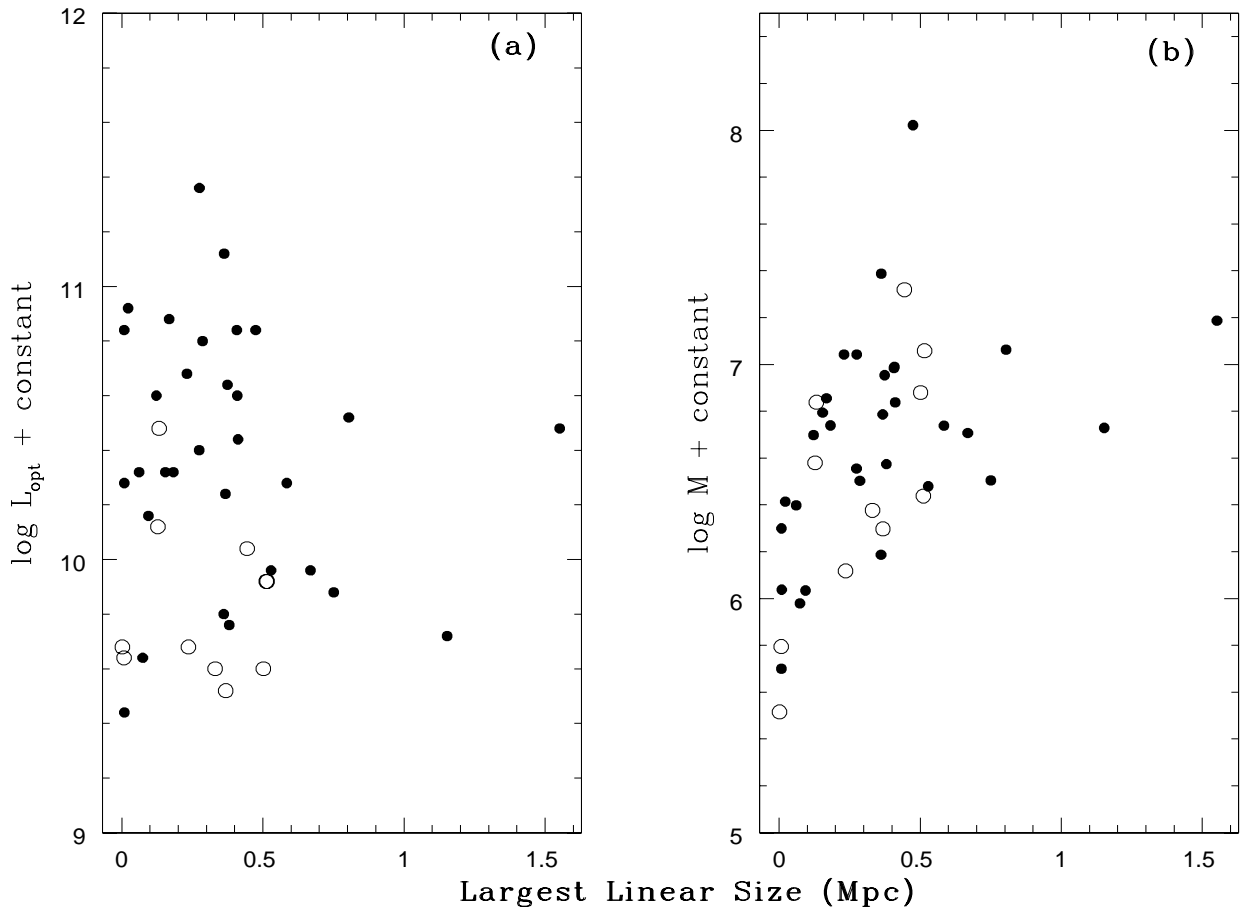}
\caption[]{{\bf (a,b):}  Plots of radio size
($l$) {\it versus} the optical luminosity and the mass of the
central engine, respectively, for our combined dataset (see Sect. 4).  The
filled circle are the data from B(96) while the open circles refer to
the data from JB(91), as explained in Sect. 2.} 
\end{figure*}

\section {Discussion}

An important inference from Fig. 1 is that over the lifetime of a typical
powerful
double radio source, the FWHM of the $H\beta$ emission line from the
BLR undergoes an increase by roughly a factor of 3--4. This broad line is
presumed to arise from the vicinity of the central super-massive
black-hole (SMBH) and its width is commonly attributed to the
gravitational influence of the SMBH, in accordance with the standard
paradigm for AGN (however, for radio-quiet AGN, alternative schemes
have been put forward by, e.g., Terlevich et al., 1992). 
In the past, widths of the
BLR emission lines have often been used as the key tracer of the mass
of the central engine of AGN (e.g., Dibai, 1984; Wandel \& Yahil, 1985;
Padovani \& Rafanelli, 1988, Perry, 1992).

Here it may be recalled that the radio spectral mapping
programmes have yielded typical lifetime of order $10^7 yr$ for
powerful double radio sources (e.g., Alexander \& Leahy, 1987; Carilli
et al., 1991; Liu \& Pooley, 1992). However, independent considerations
based on the dynamics of the radio hot-spots suggest that the true ages
are probably an order-of-magnitude larger (i.e., $\sim~10^8~yr$). This
is borne out, for instance, from a recent analysis of the lobe-length
asymmetry, in which the near-side radio lobe in individual sources could
actually be identified using the observed one-sided jet (Scheuer, 1995). 
Indeed, several authors have argued that the analyses of
spectral gradients in radio lobes could substantially underestimate the
ages, since they usually incorporate the simplistic assumption of a
uniform magnetic field. As indicated by the well-resolved maps of radio
lobes and jets, the field is likely to be concentrated in filaments
occupying just a tiny fraction of the total volume of the lobes (e.g.,
Perley, Dreher \& Cowan, 1984; Owen, Hardee \& Cornwell, 1989).
Consequently, the energetic particles would spend most of the time in the
weak-field region, thereby reducing the synchrotron losses by a large factor 
(see, e.g., Eilek, Melrose \& Walker, 1997 and references therein; 
Gopal-Krishna, 1980).  Likewise, the spectral ages may
also get underestimated if (large-scale) spatial gradients of magnetic
field are present within the radio lobes/jets (Wiita \& Gopal-krishna,
1990). Based on all these considerations, it seems more likely that the
true ages of powerful radio sources are close to $10^8 yr$. For the case 
of Eddington accretion rate one would then expect typically an 
order-of-magnitude increase in the mass of the SMBH during the 
lifetime of the radio sources, since the e-folding time for the 
black-hole mass is:
\begin{equation}
t_{Edd}=\vert \dot{M}_{Edd}/M\vert^{-1}
       =(4.5\times 10^{7}\; yr)\; (\eta/0.1),
\end{equation}
where $\eta$ is the radiative efficiency. We  argue below that even though
the persistence of an Eddington accretion rate is not supported by
the data,  the inference
about an order-of-magnitude increase in the mass of the central engine
(i.e. the region within the BLR) during the radio source lifetime
probably still holds.

In the event of Eddington accretion rate the
increase in the mass of the central black-hole with time would result
in a similar increase in the luminosity.
As a result, one would expect a positive correlation between $L_{opt}$ and $l$,
which  is not evident in  our combined data set (Fig. 2a). This weakens
considerably the case for a persistent accretion at the Eddington rate.
On the other hand, in order to estimate the increase in the mass of the central
engine one needs to take into account the result from 
the `reverberation mapping' of
nearby AGNs, which has revealed a positive dependence of the BLR radius
(r) on the luminosity($L$):  $r~\sim 0.06 L_{46}^{0.5}~pc$ (Netzer \&
Laor 1993), where $L_{46}$ is the bolometric luminosity expressed in
the units of $10^{46} erg.s^{-1}$.  Combining this with the relation
$v^2 \propto G M / r$ gives:  $M \propto v^2 \sqrt{L_{46}}$. Fig. 2b
shows a plot of $M$ {\it versus} linear size, $l$ which
is a measure of age of the quasar (note that $L$ has been approximated
by optical luminosity,  cf. Joly et al., 1985).  A positive correlation 
between $M$ and $l$ at a $4.5\sigma$ level is seen in our combined dataset of
LDQs. This suggests that over the lifetime of a typical radio quasar, 
the mass of the region inside the BLR does go up by about an 
order-of-magnitude.
These estimates could be 
further sharpened when the estimates of bolometric luminosity
for the individual quasars become available.  Another interesting 
improvement to this study would be to induct additional quasars with very 
large radio structures. That would shed light on the late 
evolutionary stages of the central engine of quasars.

{\bf Acknowledgments} It is a pleasure to thank Prof.\ Paul Wiita
and Dr. Arun Mangalam for helpful discussions. Thanks are also due
to Prof.  M. Salvati, the referee of our paper, for his insightful
comments and suggestions.

\end{document}